\documentclass[nojss]{jss}

\usepackage{lmodern}

\usepackage{pslatex}
\usepackage{textpos}
\usepackage{caption}

\usepackage{algorithm}
\usepackage{algorithmicx}
\usepackage{algpseudocode}
\usepackage{listings}
\usepackage{color}
\usepackage{textcomp}

\usepackage{graphicx}
\usepackage{epstopdf}

\usepackage{amsmath,amssymb,amsfonts,amsthm,breqn,amsbsy}
\usepackage{longtable}
\usepackage{verbatim}
\usepackage{float}
\usepackage{appendix}
\usepackage{array}
\usepackage{varwidth}
\usepackage{multirow}
\usepackage{tabularx}
\usepackage{framed}
\usepackage{lipsum}

\usepackage{indentfirst}
\usepackage{longtable}

\usepackage[utf8]{inputenc}
\usepackage{booktabs}
\usepackage{lscape}
\usepackage{rotating}
\usepackage{fancyvrb}

\usepackage{setspace}
\usepackage[T1]{fontenc}
\usepackage{indentfirst}
\usepackage{lineno}
\usepackage{etoolbox}
\usepackage{mathtools}
\usepackage{bigints}
\usepackage{url}
\usepackage{bm,bbm}
\usepackage[normalem]{ulem}

\def\htheta{\hat{\theta}}

\def\models{\mathcal{M}}

\author{Alberto Caimo\\Technological University Dublin
   \And Lampros Bouranis\\Accenture Greece
   \AND Robert Krause\\Link\"{o}ping University\\Freie Universit{\"a}t Berlin
   \And Nial Friel\\University College Dublin}
\Plainauthor{Alberto Caimo, Lampros Bouranis, Robert Krause, Nial Friel}

\title{Statistical Network Analysis with \pkg{Bergm}}
\Plaintitle{Statistical Network Analysis with Bergm}
\Shorttitle{Statistical Network Analysis with \pkg{Bergm}}

\Abstract{
Recent advances in computational methods for intractable models have made network data increasingly amenable to statistical analysis. Exponential random graph models (ERGMs) emerged as one of the main families of models capable of capturing the complex dependence structure of network data in a wide range of applied contexts. The \pkg{Bergm} package for \proglang{R} has become a popular package to carry out Bayesian parameter inference, missing data imputation, model selection and goodness-of-fit diagnostics for ERGMs. 
Over the last few years, the package has been considerably improved in terms of efficiency by adopting some of the state-of-the-art Bayesian computational methods for doubly-intractable distributions. 
Recently, version 5 of the package has been made available on CRAN having undergone a substantial makeover, which has made it more accessible and easy to use for practitioners. New functions include data augmentation procedures based on the approximate exchange algorithm for dealing with missing data, adjusted pseudo-likelihood and pseudo-posterior procedures, which allow for fast approximate inference of the ERGM parameter posterior and model evidence for networks on several thousands nodes. 
}

\Plainkeywords{Bayesian inference, exponential random graph models, R packages}

\Address{
  Alberto Caimo\\
  School of Mathematical Sciences\\
  Technological University Dublin\\
  Grangegorman D07 ADY7\\
  Dublin, Ireland\\
  E-mail: \email{alberto.caimo@tudublin.ie}\\
  URL: \url{https://acaimo.github.io}
}

\begin{document}

\section{Introduction}
Exponential random graph models (ERGMs) (\cite{fra:str86,was:pat96,rob:pat:kal:lus07}) are one of the most important family of statistical models conceived to capture the complex dependence structure of an observed network, allowing 
to identify the relational effects that are supposed to describe the link creation process.

Bayesian inference for ERGMs is challenging because of the intractability of both the likelihood and the marginal likelihood. The advanced computational methods developed by several recent papers (see for example, \cite{kos:rob:pat10,cai:fri11,cai:mir15,alq:fri:eve:bol14,bou:fri:mai17,bou:fri:mai18}) have made it possible and computationally feasible to model increasingly large network data using ERGMs on several thousands of nodes. The development of user-friendly software has always represented an essential aspect of the research activity in this area.

There is a wide range of \proglang{R} packages \citep{R} implementing various inferential approaches and modelling extensions of the ERGM framework. These include the \pkg{ergm} package \citep{ergm1} providing a comprehensive set of functions for fitting, simulating and diagnosing ERGMs; the \pkg{tergm} \citep{term} and \pkg{btergm} \citep{btergm} packages for the analysis of temporal ERGMs; \pkg{mlergm} \citep{mlergm} package for analysing multilevel ERGMs; and \pkg{fergm} package \citep{fergm} implementing estimation and fit assessment for frailty ERGMs. The main software alternative to these \proglang{R} packages is the free \pkg{PNET} programme \citep{pnet} which runs on Windows.

The \pkg{Bergm} package for \proglang{R} implements Bayesian analysis for exponential random graph models, providing a comprehensive inferential framework for Bayesian parameter estimation and model selection using efficient Monte Carlo algorithms. It can also supply model assessment and goodness-of-fit procedures that address the issue of model adequacy. Although computationally intensive, the package is easy to use and represents an attractive way of analysing networks by adopting a fully probabilistic treatment of uncertainty of the network effects that are assumed to be able to explain the overall connectivity structure of the network.
The \pkg{Bergm} package depends on the \pkg{ergm} package which is part of the \pkg{statnet} suite of packages \citep{statnet} and therefore it makes use of the same model set-up and network simulation algorithms. 
The \pkg{Bergm} package has been improved considerably in terms of usability for practitioners and performance since its early versions \citep{cai:fri14}. The aim of the package is to provide a set of tools for both developers and end-users. For this reason the package includes several competing functions (based on different statistical approaches) for accomplishing posterior parameter estimation and model selection.

The package has been used in several applications such as neuroscience \citep{sink16}, organisation science \citep{cai:lom15,tas:cai19} and political science \citep{henning19}.

The purpose of this paper is to provide a complete description of the recent improvements by summarising some of the technical background and newly implemented algorithms in the main functions of the package.

\section{Getting Bergm}
The \pkg{Bergm} package \citep{Bergm} can be obtained from CRAN and loaded in \proglang{R} using the following commands:

\begin{CodeChunk}
\begin{CodeInput}
R> install.packages("Bergm")
R> library("Bergm")
\end{CodeInput}
\end{CodeChunk}

\pkg{Bergm} depends on \pkg{ergm} \citep{hun:han:but:goo:mor08}, \pkg{network} \citep{net2,net1}, \pkg{coda} \citep{coda}, \pkg{MCMCpack} \citep{mcmcpack}, \pkg{Matrix} \citep{matrix}, \pkg{mvtnorm} \citep{mvtnorm} and \pkg{matrixcalc} \citep{matrixcalc}. Loading the package will automatically load all the dependencies. All of these packages are available on the Comprehensive R Archive Network (CRAN) at \url{http://CRAN.R-project.org/}. The results presented in this paper have been obtained using \proglang{R} version 4.0.2; \pkg{Bergm} version 5.0.2; \pkg{ergm} version 3.11; \pkg{network} version 1.16.1; \pkg{coda} version 0.19-4; \pkg{MCMCpack} version 1.4-9; \pkg{Matrix} version 1.2-18; \pkg{mvtnorm} version 1.1-1 and \pkg{matrixcalc} version 1.0-3.

\section{Network data}

Two very well known network datasets are used throughout this tutorial for illustrative purposes: 
the first is the Lazega's Law Firm dataset whose undirected edges represent collaborative relations in a Northeastern US corporate law firm \citep{laz01}; the second is the Faux Dixon High School dataset, which represents a directed friendship network \citep{resnick}.

\subsection{Lazega's law firm}

\begin{figure}
\centering
\includegraphics[width=15cm]{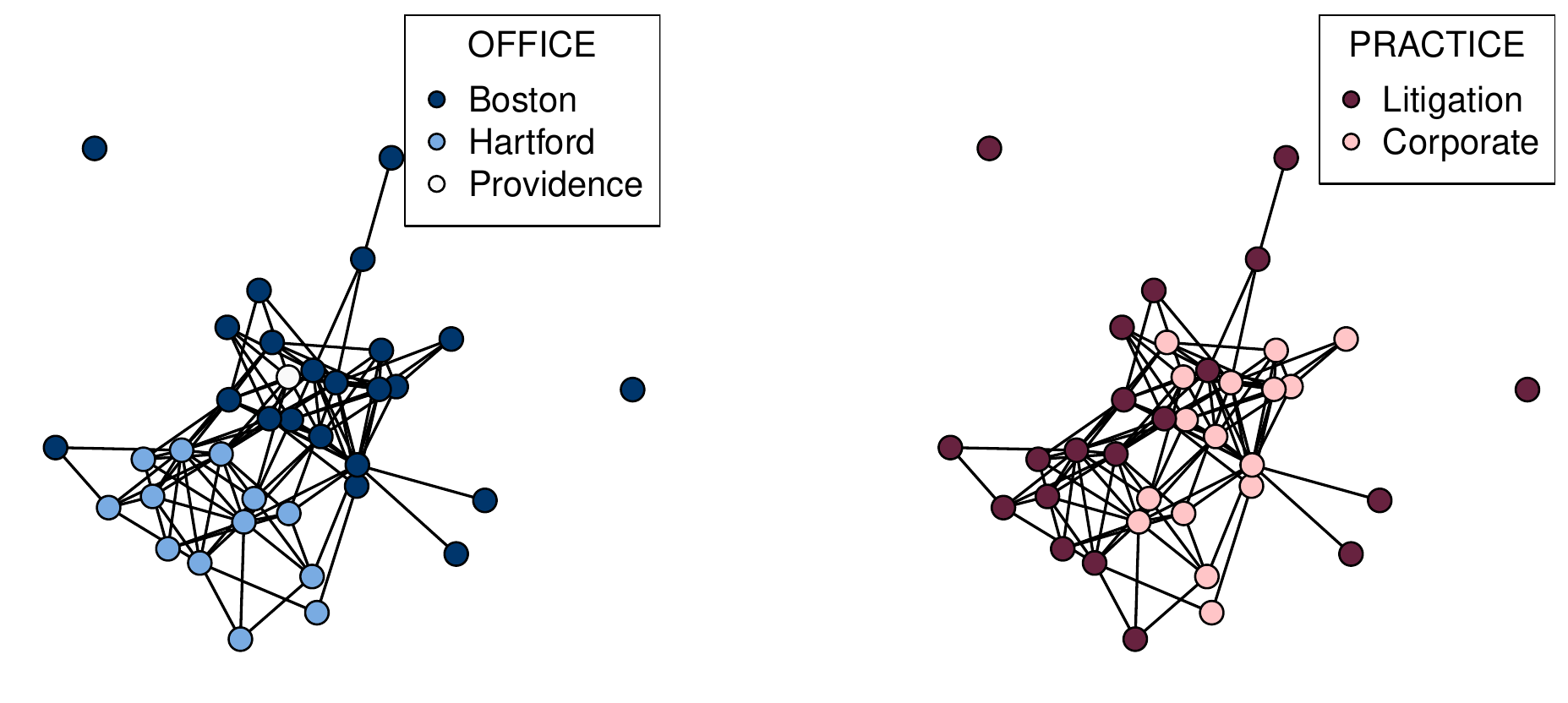}
\caption{Lazega's network undirected graphs.}
\label{fig:lazega}
\end{figure}

The law office in Lazega's study network (graphs displayed in Figure~\ref{fig:lazega}), consists of binary undirected collaborative relations between 36 partners in a Northeastern US corporate law firm \citep{laz01}. Various members' attributes are also part of the dataset, including seniority, formal status, office in which they work, gender, law school attended, individual performance measurements (hours worked, fees brought in), attitudes concerning various management policy options.

\begin{CodeChunk} 
\begin{CodeInput} 
R> data(lazega)
\end{CodeInput} 
\end{CodeChunk} 

\subsection{Faux Dixon High School}
\begin{figure}
\centering
\includegraphics[width=15.5cm]{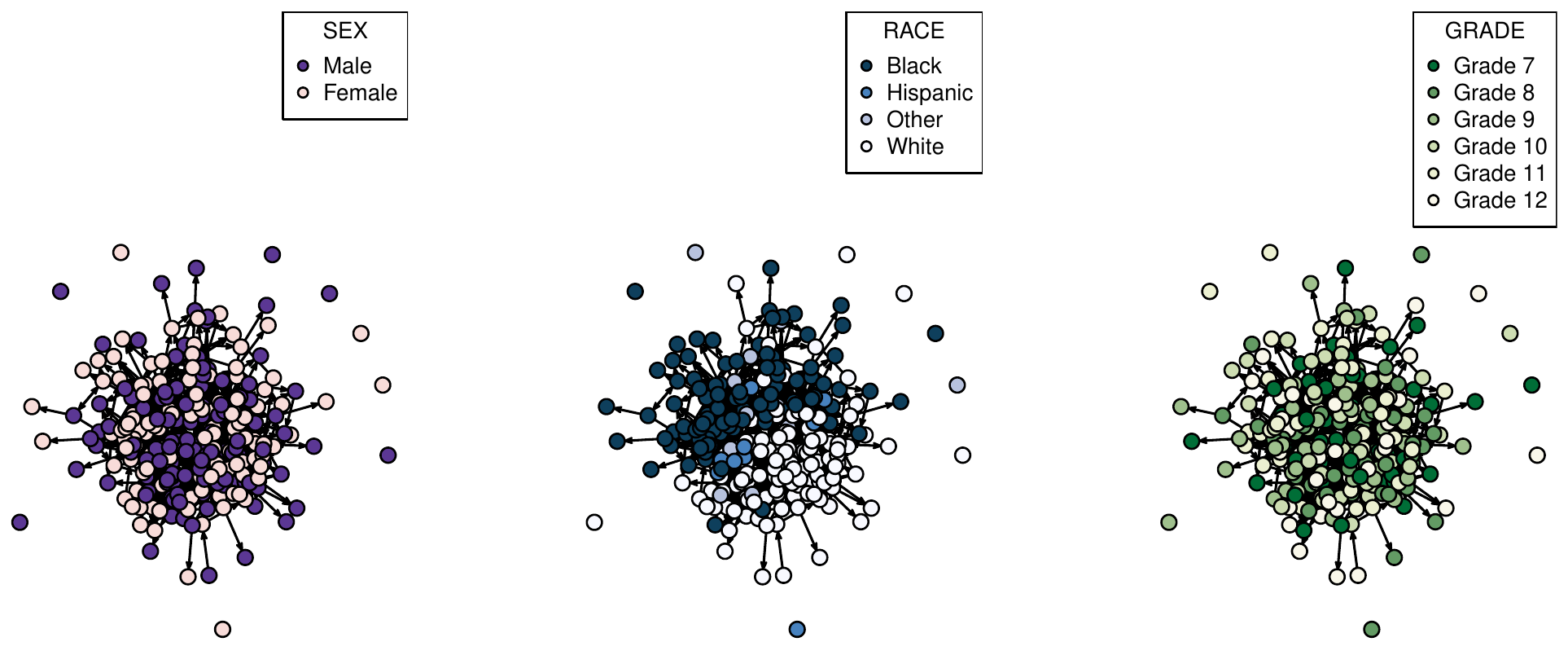}
\caption{Faux Dixon High School network directed graphs.}
\label{fig:dixon}
\end{figure}

The Faux Dixon High School network data (graphs displayed in Figure~\ref{fig:dixon}), included in the \pkg{ergm} package, represents a simulation of a binary directed in-school friendship network \citep{resnick}. 
See \code{?faux.dixon.high} for the ERGM that was fit to the original data, generating the network dataset.
The network comprises 248 nodes representing students. Information on the following nodal attribute variables is available: sex, race, grade.

\begin{CodeChunk} 
\begin{CodeInput} 
R> data(faux.dixon.high)
R> dixon <- faux.dixon.high
\end{CodeInput} 
\end{CodeChunk} 

The code used to plot the network graphs in Figures~\ref{fig:lazega} and \ref{fig:dixon} is included in the R replication code of the supplementary files.

\section{Exponential random graph models}\label{sec:ergm}
Networks are relational data defined as a collection of actors interacting with each other and connected in a pairwise fashion. Networks can be represented as graphs consisting of a set of $n$ nodes and a set of edges which define some sort of relationships between pairs of nodes (dyads).
The connectivity pattern of a graph can be described by an $n \times n$ adjacency matrix $Y$ encoding the presence or absence of an edge between nodes $i$ and $j$:
\begin{equation*}
  Y_{ij}=\begin{cases}
	        1, &\textrm{if }i\textrm{ and }j \textrm{ are connected,}\\
            0, &\textrm{otherwise.}\\
          \end{cases}
\end{equation*}
If the network is undirected, then $y_{ij} = y_{ji}$ and the adjacency matrix is symmetric, otherwise the network is directed.
Edges connecting a node to itself (self-loops) are generally not allowed in many applications and will not be considered in this context.

Introduced by \cite{hol:lei81} to model individual heterogeneity of nodes and reciprocity of their edges, the family of exponential random graph models (ERGMs) was generalised by \cite{fra:str86}, \cite{was:pat96} and \cite{sni:pat:rob:han06} in order to account for higher-order dyadic relationships.
ERGMs constitute a broad class of network models (see \cite{rob:pat:kal:lus07} for an introduction) assuming that the probability of an observed network $y$ can be explained in terms of the relative prevalence of a set of network statistics $s(y)$:
\begin{equation}\label{eqn:ergmprob}
f(y\mid\theta)=\frac{\exp\left\{\theta^{\top}s(y)\right\}}{z(\theta)},
\end{equation}
where $\theta\in\Theta\subseteq\mathbb{R}^{d}$ is the vector of $d$ model parameters associated with $s(y)$ \citep{hunter,sni:pat:rob:han06}, $z(\theta) = \sum_{y\in \mathcal{Y}}^{} \exp\left\{\theta^{\top}s(y)\right\}$ is the likelihood normalising constant consisting of a sum across $\mathcal{Y}$ which is the set of all possible graphs on $n$ nodes. Additionally, ERGMs allow for incorporation of covariate information $X$, e.g., the number of network statistic configurations within the same attribute category. 

The normalising constant $z(\theta)$ is computationally tractable only for trivially-small networks, as $\mathcal{Y}$ involves $2^{\binom{n}{2}}$ and $2^{n(n-1)}$ possible undirected and directed graph configurations respectively. 
The network statistics $s(y)$ often represent a series of counts of sub-graph configurations (e.g., the number of edges, stars, triangles, functions of degree distributions, edgewise shared partners), that capture the relevant information of the global connectivity structure of the network graph \citep{sni:pat:rob:han06}. 

Given~\eqref{eqn:ergmprob} we can express the distribution of the Bernoulli variable $Y_{ij}$ under the conditional form
\begin{equation*}
\text{logit}\left\{\Pr(Y_{ij} = 1 \mid y_{-ij},\theta)\right\}=\theta^{\top}\delta_{s}(y)_{ij},
\end{equation*}
where $\delta_{s}(y)_{ij}=s(y^{+}_{ij})-s(y^{-}_{ij})$ denotes the vector representing the change in the vector of network sufficient statistics when the value of $y_{ij}$ is toggled from a 0 (empty dyad, $y^{-}_{ij}$) to a 1 (edge, $y^{+}_{ij}$), holding the rest of the network fixed: $y_{-ij}=y\setminus\{y_{ij}\}.$ \cite{strauss} applied the pseudo-likelihood method of \cite{besag2} to social networks which aims to approximate the full joint distribution in \eqref{eqn:ergmprob} by the product of the full conditional probabilities of the network dyads:
\begin{equation}\label{eq:pseudolike}
f_{\text{PL}}(y\mid \theta) 
= \prod_{\substack{i\neq j\ \cup\ i< j}} p(y_{ij} \mid y_{-ij},\theta),
\end{equation}
where the conditions $i \ne j$ and $i < j$ hold for directed and undirected networks, respectively.
The pseudo-likelihood is equivalent to a logistic regression model \citep{was:pat96} and ignores strong dependencies that might exist among network edges (e.g., transitivity effect) in the data: it can therefore lead to a biased estimation. 

\section{Parameter estimation}\label{sec:bergm}
Bayesian analysis is a fully probabilistic treatment of uncertainty related to models and model parameters. A major advantage of the Bayesian approach is the flexibility with which prior information on the uncertainties related to the models and their parameters can be incorporated. In fact, the Bayesian approach permits the researcher to use both data and prior (e.g., expert-judgement) information in a consistent manner. For example, prior information about the data and/or from previous studies can easily be incorporated through an informative prior distribution (see for example \cite{cai:pal:lom17,baletal19}). This can be done by simply placing prior probability distributions on the possible values of the unknown parameters or models. 

Let $p(\theta)$ be the prior distribution for the model parameters, $\theta$. The posterior distribution of the parameters given the data can be obtained by using the Bayes' theorem:
$$
\pi(\theta \mid y) = \frac{f(y \mid \theta)\ p(\theta)}{\pi(y)},
$$
where $\pi(y) = \int_{\Theta} f(y\mid \theta)p(\theta)\;\mathrm{d}\theta$ is the normalising function for the posterior distribution, termed marginal likelihood or model evidence. In the ERGM context, both $z(\theta)$ and $\pi(y)$ are typically intractable and the posterior distribution is therefore computationally doubly-intractable. 

\subsection{The approximate exchange algorithm}

In order to approximate the posterior distribution $\pi(\theta \mid y)$, the \pkg{Bergm} package uses the exchange algorithm described in Section 4.1 of \cite{cai:fri11} to sample from the following distribution:
\begin{equation*}
\pi(\theta',y',\theta \mid y) \propto f(y \mid \theta)\ p(\theta)\ \epsilon(\theta' \mid \theta)\ f(y'\mid \theta') 
\end{equation*}
where $f(y' \mid \theta')$ is the likelihood on which the simulated data $y'$ are defined and belongs to the same exponential family of densities as $f(y \mid \theta)$, $\epsilon(\theta' \mid \theta)$ is any arbitrary proposal distribution for the augmented variable $\theta'$.
At each MCMC iteration, the exchange algorithm consists of a Gibbs update of $\theta'$ followed by a Gibbs update of $y'$, which is drawn from the $f(\cdot \mid \theta')$ via an MCMC algorithm \citep{hun:han:but:goo:mor08}. Then an exchange or swap from the current state $\theta$ to the proposed new parameter $\theta'$ is performed. 

\subsubsection[Parallel adaptive direction sampler]{Parallel adaptive direction sampler}
\label{sec:ads}

In order to improve mixing a parallel adaptive direction sampler (ADS) \citep{gil:rob:geo94,rob:gil94} is considered: at the $i$-th iteration of the algorithm we have a collection of $H$ different chains interacting with one another. By construction, the state space consists of $\{\theta_1,\dots,\theta_H\}$ with target distribution $\pi(\theta_1|y)\otimes\dots\otimes \pi(\theta_H | y)$. A parallel ADS move consists of generating a new parameter value $\theta'_h$ from the difference of two parameters $\theta_{h_1}$ and $\theta_{h_2}$ (randomly selected from other chains) multiplied by a scalar term $\gamma$ (called ADS move factor) plus a random proposal $\epsilon$ (which is a multivariate Normal distribution in \code{bergm()}). 

The format of the model specification is the same of the \pkg{ergm} package formula (use \code{?ergm-terms} to get a complete list of network statistics implemented in the \pkg{ergm} package). Let us consider the Lazega's law Firm network and a model including the following network statistics:
\begin{CodeChunk} 
\begin{CodeInput} 
R> m1 <- lazega ~ edges +
+    nodematch("Office") +
+    nodematch("Practice") +
+    gwesp(0.5, fixed = TRUE)
\end{CodeInput} 
\end{CodeChunk} 

In this case, our focus is on the density effect captured by the number of edges (\code{edges}), the homophily effect between lawyers working in the same office (\code{nodematch("Office")}) and in the same practice area (\code{nodematch("Practice")}), and the transitivity effect captured by the geometrically weighted edgewise shared partners statistic (GWESP) with fixed decay parameter equal to 0.5 (\code{gwesp(0.5, fixed = TRUE)}) \citep{sni:pat:rob:han06}.

As mentioned in Section~\ref{sec:bergm}, we can specify prior distributions for the parameters in the model. The \code{bergm()} function allows users to specify the mean vector and variance/covariance matrix of a multivariate Normal distribution. For example we can create the \code{M.prior} object where we set the mean for the first parameter (corresponding to the \code{edge} statistic) to be equal to -4 which corresponds to assuming a priori that the average conditional odds of an edge between any two nodes $i$ and $j$ is $\exp(-4) \approx 0.018.$ This reflects our prior assumption of overall sparsity of the network. The other prior mean values are set to be positive reflecting our prior assumption of positive homophily effect and transitivity as generally observed in this kind of social networks. The \code{S.prior} object is set to be a diagonal matrix with variances equal to 4.

\begin{CodeChunk} 
\begin{CodeInput} 
R> M.prior <- c(-4, 0.5, 0.5, 1)
R> S.prior <- diag(4, 4)
\end{CodeInput} 
\end{CodeChunk} 

By adopting the parallel ADS procedure we need to set the number of parallel chains by using the argument \code{nchains}. The number of chains must be greater than 3 and it is by default set to be twice the model dimension. For each chain, we can then set the number of burn-in iterations (\code{burn.in}) and the number of iterations after the burn-in (\code{main.iters}). The number of iterations used to simulate a network $y'$ at each iteration is defined by the argument \code{aux.iters}. 

The arguments \code{prior.mean} and \code{prior.sigma} allow us to specify the multivariate Normal parameter prior distribution defined above.

The total number of iterations, i.e., the size of the posterior sample, is \code{nchains} $\times$ \code{main.iters}. The proposal covariance structure of the proposal distribution $\epsilon(\cdot)$ is defined by the argument \code{V.proposal} which by default is set to be a diagonal matrix with every diagonal entry equal to 0.0025. In many cases, good mixing of the chain is ensured by a sensible tuning of the parallel ADS move factor \code{gamma} and therefore the argument \code{V.proposal} can be generally left at its default value. The parameter \code{gamma} can be easily tuned to achieve a suitable acceptance rate ($\sim 20\%$) by starting from its default value ($0.5$): the higher the value of \code{gamma} the lower the acceptance rate and vice versa. The range of values that \code{gamma} can take depends on the size of network and the kind of network statistics included in the model.
The \pkg{Bergm} functions can only estimate ERGMs with dimensions greater that 1, meaning that at least 2 network statistics must be included in the model specification.

\begin{CodeChunk} 
\begin{CodeInput} 
R> p.m1 <- bergm(m1,
+    prior.mean  = M.prior,
+    prior.sigma = S.prior,
+    burn.in     = 500,
+    main.iters  = 3000,
+    aux.iters   = 2500,
+    nchains     = 8,
+    gamma       = 0.6)
\end{CodeInput} 
\end{CodeChunk}

It is possible to summarise the posterior results of the MCMC estimation procedure by using the \code{summary()} function.

\begin{CodeChunk} 
\begin{CodeInput} 
R> summary(p.m1)
\end{CodeInput} 
\end{CodeChunk} 

\begin{CodeChunk} 
\begin{CodeOutput}
 Posterior Density Estimate for Model: y ~ edges + 
    nodematch("Office") + nodematch("Practice") + gwesp(0.5, fixed = TRUE) 
 
                                  Mean        SD    Naive SE Time-series SE
theta1 (edges)              -5.1103044 0.4508997 0.002910545    0.019564602
theta2 (nodematch.Office)    0.9257179 0.1813198 0.001170414    0.007285688
theta3 (nodematch.Practice)  0.6454322 0.1861145 0.001201364    0.007621859
theta4 (gwesp.fixed.0.5)     1.5173614 0.2514910 0.001623368    0.010569240

                                  2.5%        25%        50%        75%     97.5%
theta1 (edges)              -6.0221370 -5.4173036 -5.0952047 -4.7993651 -4.251420
theta2 (nodematch.Office)    0.5773097  0.8016098  0.9223223  1.0464544  1.278189
theta3 (nodematch.Practice)  0.2763161  0.5215024  0.6443654  0.7703229  1.006394
theta4 (gwesp.fixed.0.5)     1.0414392  1.3449247  1.5108637  1.6893190  2.026859

 Acceptance rate: 0.2 
\end{CodeOutput}
\end{CodeChunk}

The output above shows the results of the MCMC estimation: posterior means, standard deviations, medians, posterior quantile values and overall acceptance rate. In this example the $95\%$ credible interval for the \code{edges} parameter $\theta_1$ lies in the negative region whereas the $95\%$ credible intervals for all the other parameters $\theta_2, \theta_3, \theta_4$ lie in positive regions. This means that the baseline edge probability is low and most of the edges of the network tend to connect nodes within the same office and in the same practice area and they tend to form triadic relations.

\begin{figure}[H]
\centering
\includegraphics[width=15cm]{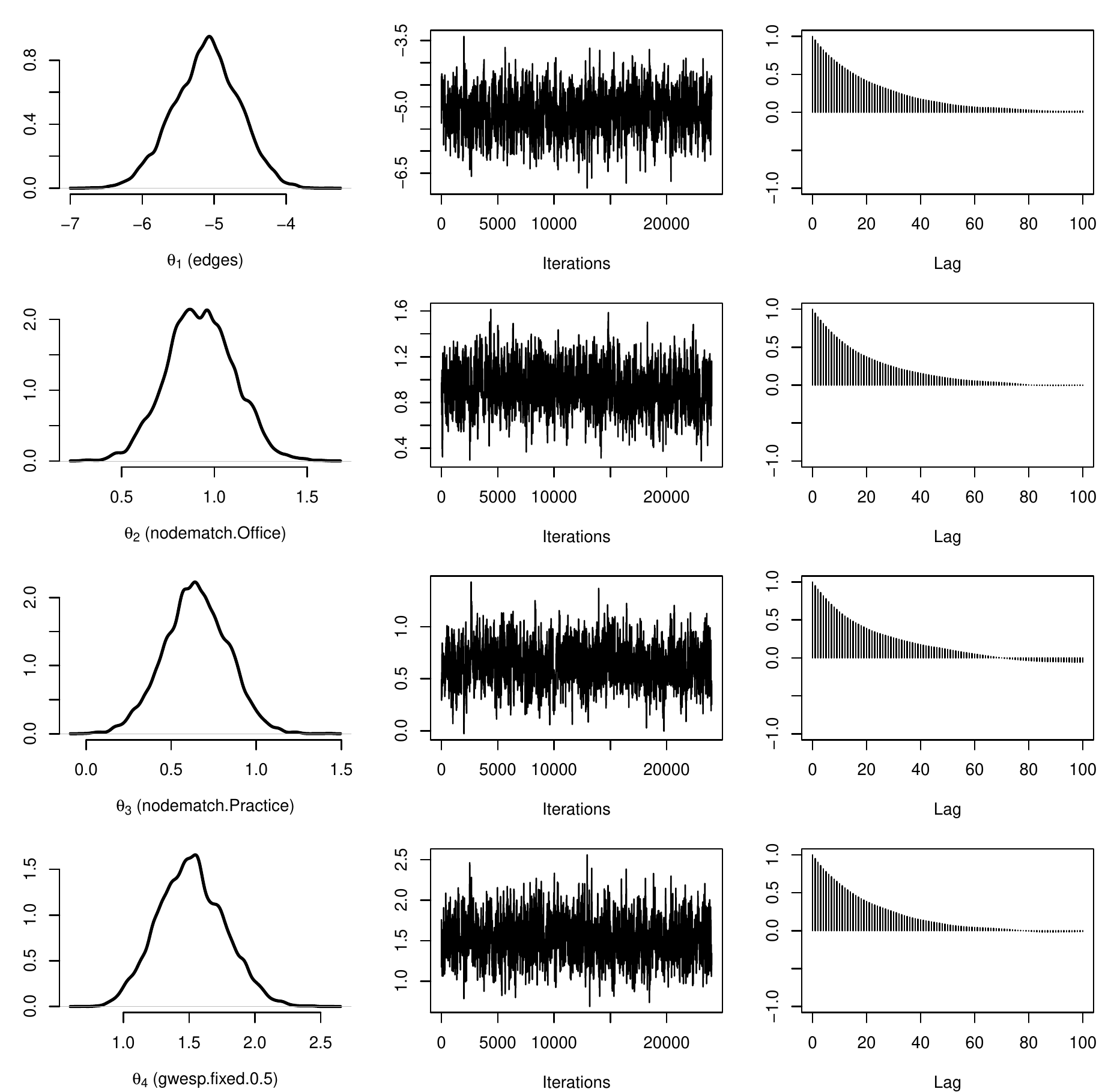}
\caption{MCMC diagnostics for model \code{p.m1}.}
\label{fig:p_m1_traces}
\end{figure}

Figure~\ref{fig:p_m1_traces} displays the MCMC diagnostic plots produced by the \code{plot()} function. The argument \code{lag} allows us to set the maximum lag for which autocorrelation is computed. 
The overall acceptance rate is $19\%$ and the autocorrelation is negligible after lag 60; the traceplots indicate good mixing of the MCMC algorithm.

\begin{CodeChunk} 
\begin{CodeInput} 
R> plot(p.m1, lag = 100)
\end{CodeInput} 
\end{CodeChunk} 

\subsection{Missing data augmentation}
The estimation algorithm described above can lead to biases if $y$ is not fully observed, because $s(y)$ cannot be properly calculated. A solution for this problem was proposed by \citet{kos:rob:pat10} and has been evaluated by \citet{kraMiss20}. Let $I$ be an indicator matrix of whether a tie variable is observed or missing, with $I_{ij} = 1$ if $y_{ij}$ is observed and $I_{ij} = 0$ if $y_{ij}$ is missing. Further we use the convention that $u$ represents the observed part of the data ($I_{ij} = 1$) and $v$ represents the unobserved part of the data ($I_{ij} = 0$). Thus the network can be reassembled as $y = (u,v)$. With the given network we can define an observation model for $I$, $f(I \mid y,\zeta)$, which is a probability model for what is observed and what is not, depending on the network $y$ and some statistical parameter $\zeta$. We present the algorithm limited to the setting where $I$ is known and fixed and all covariates are known and fixed; extensions for multiplex networks exist \citep{kra:cai19}. Further, the algorithm assumes that the missing data is ignorable, meaning the probability for data to be missing is independent from the missing values themselves and only dependent on the observed data. In other words, observations are "missing at random" (or "missing completely at random") \cite{ru87}. Thus we can ignore the missing data mechanism $\zeta$ in the estimation process. We now use data augmentation to estimate $\theta$ under missing data by alternating between draws from $\theta \mid u,v$ and $v \mid u,\theta$. We augment the observed data $u$ by draws from the unobserved data $v$ from the full conditional posterior $v^* \leftarrow v \mid u,\theta$, creating the augmented network $y^* = (u,v^*)$.

This process is implemented in \code{bergmM()} by adapting the approximate exchange algorithm presented above in the following way: an additional step to the MCMC procedure at each iteration is included if the proposed parameter vector $\theta'$ is accepted. In this additional step a new draw from $v^* \leftarrow v \mid u,I,\theta'$ is obtained and the augmented network $y^*$ is updated. After which $y^*$ is fed back to the algorithm as starting point for the next iteration as new baseline network. The augmented network is used to obtain the sufficient statistics $s(y^*)$, thus the third step of the algorithm does not optimise $s(y') - s(y)$, but $s(y') - s(y^*)$. A naive imputation is used for the first iteration of the algorithm to obtain an augmented starting network. In first step we substitute $s(y^*)$ with $s(u)$. 

The algorithm can be summarised as follows:

\begin{algorithmic}
\State Use a naive imputation to initialise $s(y^*)$
\State Initialise $\theta$
\For {$k = 1, \dots, K$}
    \State Generate $\theta'$ using the ADS proposal procedure
    \State Simulate $y'$ from $f(\cdot \mid \theta')$
    \State Update $\theta$ $\rightarrow$ $\theta'$ with the log of the probability:\\
	\begin{equation}
	\log \alpha = \min \left(0, [\theta - \theta']^{\top}[s(y') - s(y^*)] + \log\left[\frac{p(\theta')}{p(\theta)} \right] \right)
    \end{equation}
    \If{$\theta'$ accepted} Simulate $v^*$ from $f(\cdot \mid \theta', y^*)$ and generate a new $y^* = (u,v^*)$
    \EndIf
\EndFor
\end{algorithmic}

The function \code{bergmM()} is similar to \code{bergm()} with two additional arguments, \code{nImp} and \code{missingUpdate}. The argument \code{nImp} can be used to retain a specified number of imputed networks $y^*$ from the estimation procedure. By default, imputed networks will not be returned. If more than two networks are to be retained during the estimation (\code{nImp > 2}), \code{bergmM()} will automatically space the sampling of $y^*$ maximally over all iterations in \code{main.iters}. The second argument, \code{missingUpdate}, specifies how many tie swaps of the missing tie variables are simulated to obtain $v^*$ from $f(\cdot \mid \theta', y^*)$ for each update of $y^*$. By default this is set to the number of missing edges, \code{missingUpdate = sum(is.na(y))}.

Below we illustrate the use of \code{bergmM()} on the Lazega's law firm network. Lazega's data is fully observed, thus we randomly set all outgoing ties of 4 nodes (11\%) to missing.

\begin{CodeChunk} 
\begin{CodeInput} 
R> set.seed(1)
R> missV <- sample(1:36, 4)
R> lazega[missV, ] <- lazega[, missV] <- NA

R> set.seed(1)
R> p.m1.M <- bergmM(m1,
+    prior.mean  = M.prior,
+    prior.sigma = S.prior,
+    burn.in     = 200,
+    main.iters  = 3000,
+    aux.iters   = 3000,
+    nchains     = 8,
+    gamma       = 0.6,
+    nImp        = 10)   
\end{CodeInput} 
\end{CodeChunk} 

The object returned by \code{bergmM()} is a \code{bergm} object, thus the regular functions for assessing the estimation, obtaining MCMC diagnostics, and goodness of fit (see Section~\ref{sec:gof}) apply.

\begin{CodeChunk}
\begin{CodeInput}
R> summary(p.m1.M)
\end{CodeInput}
\end{CodeChunk} 

\begin{CodeChunk} 
\begin{CodeOutput} 
 Posterior Density Estimate for Model: y ~ edges + nodematch("Office") + 
   nodematch("Practice") + gwesp(0.5, fixed = TRUE) 
 
                                  Mean        SD    Naive SE Time-series SE
theta1 (edges)              -4.7816252 0.4379100 0.002826697    0.019273461
theta2 (nodematch.Office)    0.8600249 0.1882613 0.001215222    0.008208019
theta3 (nodematch.Practice)  0.5673884 0.1950185 0.001258839    0.008632582
theta4 (gwesp.fixed.0.5)     1.3686299 0.2477281 0.001599078    0.010902701

                                  2.5%        25%        50%        75%      97.5%
theta1 (edges)              -5.7221102 -5.0601098 -4.7666853 -4.4889519 -3.9525110
theta2 (nodematch.Office)    0.4848832  0.7358885  0.8636319  0.9799519  1.2346474
theta3 (nodematch.Practice)  0.1759360  0.4410472  0.5689183  0.6984204  0.9546846
theta4 (gwesp.fixed.0.5)     0.9183192  1.1988890  1.3552242  1.5238552  1.8846317

 Acceptance rate: 0.19 
\end{CodeOutput}
\end{CodeChunk}

In this example, the estimated parameter posterior density summaries obtained by the missing data estimation procedure are consistent with the ones obtained in Section~\ref{sec:ads} with no missing data. It is possible to get the imputed networks by typing:
\begin{CodeChunk}
\begin{CodeInput}
R> p.m1.M$impNets
\end{CodeInput}
\end{CodeChunk}

\subsection{Fixing parameters during estimation}

Sometimes it is necessary for theoretical or practical reasons to fix one or more of the model parameters during estimation. For instance, \citet{kri11} have shown that using a fixed offset term during ERGM estimation can reduce the differences in parameter estimates which can result from networks having different sizes. Similar to the \pkg{ergm} package \citep{ergm1}, parameters that should be fixed during estimation need to be given as \code{offset()} in the model and a vector with values to which these parameters are to be fixed needs to be provided with the \code{offset.coef} argument. Note that one still needs to provide prior values for the fixed parameters. These values are, however, ignored during the estimation. Further, neither positive (\code{Inf}) nor negative infinity (-\code{Inf}) are valid inputs for \code{offset.coef}, instead one should use extreme numeric values (e.g., for the \code{mutual} parameter values of $100$ or $-100$ can be considered as extreme). This feature is currently only implemented for the \code{bergm()} and \code{bergmM()} functions.

\section{Goodness-of-fit diagnostics}\label{sec:gof}

\cite{hun:goo:han08} proposed systematic simulation-based goodness of fit (GOF) diagnostics for ERGMs, comparing several high-level statistics of observed networks with those of corresponding networks simulated from the estimated network. In the Bayesian framework, in order to evaluate the model goodness of fit in terms of posterior predictive assessment, the observed network is compared to a set of networks simulated from the estimated posterior distribution of the parameters of the model \cite{cai:fri11}. 

The \code{bgof()} function is used to carry out the Bayesian goodness-of-fit diagnostic procedure. The observed network is compared with a randomly simulated network sample (which size determined by the argument \code{sample.size}) drawn from the estimated posterior distribution using \code{aux.iters} iterations for the network simulation step. 

\begin{figure}[H]
\centering
\includegraphics[width=15cm]{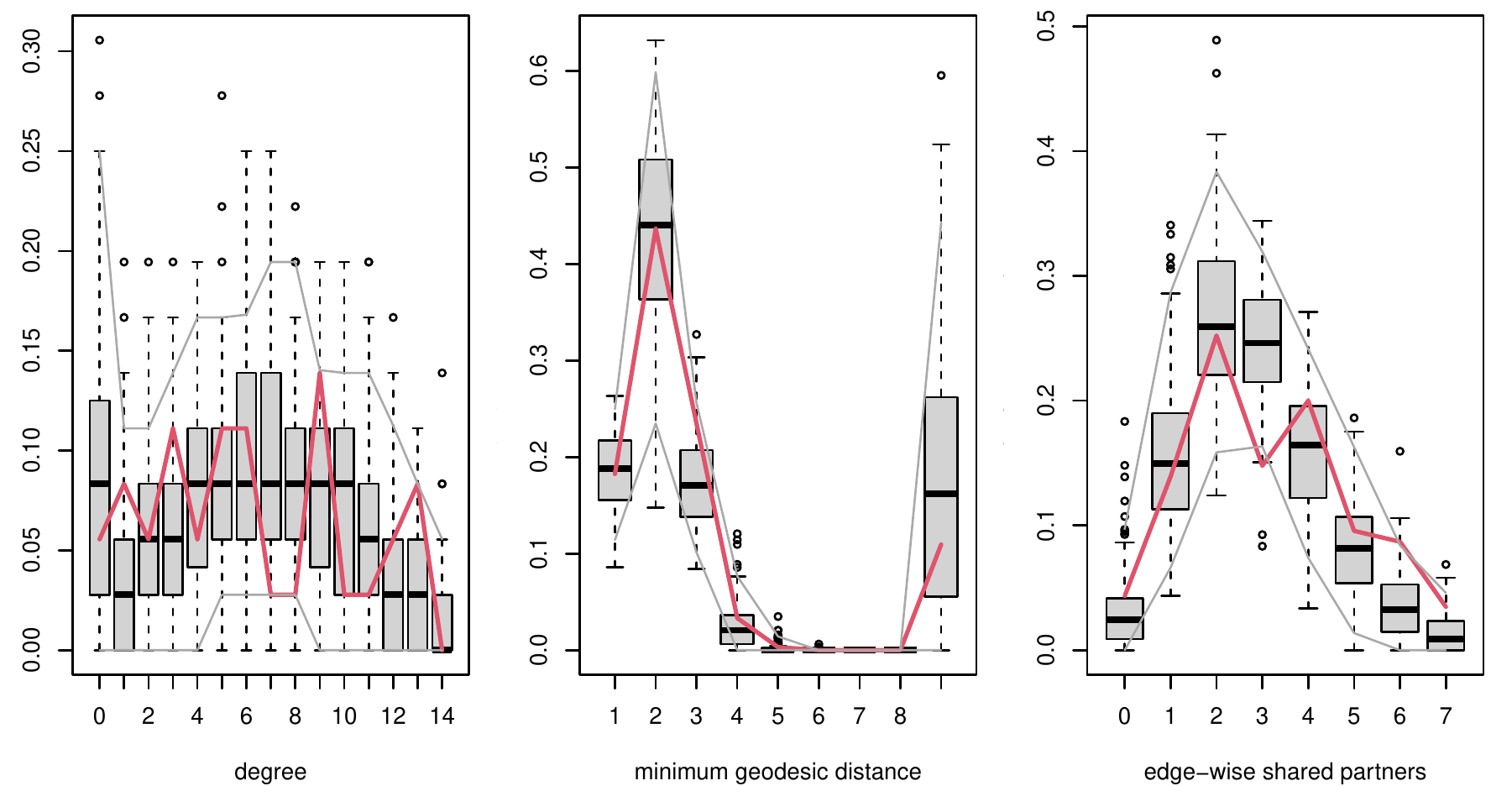}
\caption{Bayesian goodness-of-fit diagnostics for the estimated parameter posterior distribution of model \code{p.m1}.}
\label{fig:lazega_gof}
\end{figure}

In Figures~\ref{fig:lazega_gof} the red lines represent the observed network GOF statistic values, the boxplots represent the GOF statistics of the simulated networks.

The \code{bgof()}  function will produce the GOF diagnostic plots according to the type of network observed (directed or undirected). Depending on the type of network the user can specify the maximum number of GOF distributions to be plotted. The set of statistics used for the comparison of directed graphs (such as the Lazega network) includes the degree distribution, the minimum geodesic distance distribution and the edgewise shared partner distribution. The arguments \code{n.deg}, \code{n.dist}, and \code{n.esp} indicate the number of boxplots to plot starting from the minimum value for each distribution, respectively.
\begin{CodeChunk} 
\begin{CodeInput} 
R> set.seed(1)
R> bgof(p.m1,
+    aux.iters = 5000,
+    n.deg     = 15, 
+    n.dist    = 9, 
+    n.esp     = 8)
\end{CodeInput} 
\end{CodeChunk}

Figure~\ref{fig:lazega_gof} shows that the networks simulated from the estimated posterior distributions are in reasonable agreement with the observed network as the red line is almost always falling inside the $95\%$ interval represented by the light grey lines in the GOF plots.

\section{Model selection}\label{sec:bms}

A model selection (or a model choice, or a model comparison) problem is faced when several competing statistical models are considered, any of which could serve as an explanation for our data, and we would like to select the best of them. 
In the ERGM context, this task translates into the choice of which subset of network statistics should be included into the model \citep{cai:fri13}.
Let us assume a countable model set $\models=\{\models_1,\models_2,\models_3,\ldots\}$. 
Each model indexed by $m$ is defined by a likelihood function $f(y\mid\theta_m,\models_m)=f_m(y\mid\theta_m)$ and a prior on the model-specific parameter vector $\theta_m\in\Theta_m\subseteq\mathbb{R}^{d_m}$ denoted as $p(\theta_m\mid\models_m)=p_m(\theta_m)$, where  $d_m$ is the dimension of the parameter space $\Theta_m$. 
The prior beliefs for each model are expressed through a prior distribution $p(\models_m)$, such that $\sum_{m=1}^{|\models|}p(\models_m)=1$ where $|\models|$ is the cardinality of the model set.

Pairwise model comparisons can be performed with the posterior odds ratio between two models $\models_m$ and 
$\models_{m'}$:
\begin{equation}\label{eqn:por}
\frac{\pi(\models_m\mid y)}{\pi(\models_{m'}\mid y)}=
\frac{\pi(y\mid \models_m)}{\pi(y\mid \models_{m'})}\times
\frac{p(\models_m)}{p(\models_{m'})},
\end{equation}
where$\dfrac{\pi(y\mid \models_m)}{\pi(y\mid \models_{m'})}$ is the Bayes Factor; 
\begin{equation*}
\pi(\models_m\mid y) = \frac{\pi(y\mid \models_m)p(\models_m)}{\sum_{j=1}^{|\models|} \pi(y\mid \models_j)p(\models_j)}
\end{equation*}
is the posterior model probability for model $\models_m;$ and 
\begin{equation}\label{eq:evidence}
\pi(y\mid \models_m) = \int_{\Theta_m} f_m(y\mid\theta_m)p_m(\theta_m)\,\mathrm{d}\theta_m
\end{equation}
is the model evidence or marginal likelihood under model $\models_m$. 

The \pkg{Bergm} package assumes a multivariate Normal prior $\mathcal{N}_{d_m}\left(\mu_m,\Sigma_m\right)$ for $\theta_m,$ that leads to a marginal likelihood which is finite.
Only for some elementary cases the integration in \eqref{eq:evidence} can be evaluated analytically. For moderately high-dimensional problems \eqref{eq:evidence} is usually computationally intractable, and therefore sophisticated computational methods based on simulation are usually used to estimate it. There is a large number of approaches based solely on within-model simulation that provides marginal likelihood estimates by utilising the posterior samples of separate models. Recent reviews comparing popular methods based on MCMC sampling (all of which assume a tractable likelihood) can be found in \cite{wyse_review} and in \cite{ardia_review}. 

\section{Pseudo-likelihood adjustment for large networks}

Motivated by the inefficiency of the correction procedure of \cite{bou:fri:mai17} for model comparison (see \code{?bergmC} for more details), \cite{bou:fri:mai18} presented novel methodology for adjusting the pseudo-likelihood function directly as a means to obtain a reasonable and tractable approximation to the likelihood.
These adjustments involve a correction of the mode, the curvature and the magnitude at the mode of the pseudo-likelihood~\eqref{eq:pseudolike} and are implemented in the \code{ergmAPL()} function. In this function, the model-specific fully adjusted pseudo-likelihood 
\begin{equation}\label{eq:fully_adj}
\tilde{f}_m(y\mid \theta_m)=C_m\cdot f^{}_{PL,m}(y\mid \htheta_{MPLE,m}+Q_m(\theta_m-\htheta_{MLE,m})),
\end{equation}
depends on the maximum likelihood estimate, $\htheta_{MLE,m}$, the maximum pseudo-likelihood estimate, $\htheta_{MPLE,m}$, an upper triangular matrix $Q_m$ of order $d_m$ and the magnitude adjustment constant $C_m>0$. 
Most crucially, \eqref{eq:fully_adj} renders the corresponding posterior distribution 
\begin{equation}\label{eq:cor_pp}
\widetilde{\pi}(\theta\mid y,\models_m)=
\frac{\tilde{f}(y\mid \theta_m,\models_m)p(\theta_m\mid\models_m)}{\widetilde{\pi}(y\mid \models_m)}=
\frac{\tilde{f}(y\mid \theta_m,\models_m)p(\theta_m\mid\models_m)}{\int_{\Theta_m} \tilde{f}(y\mid \theta_m,\models_m)p(\theta_m\mid\models_m)\;\mathrm{d}\theta_m}
\end{equation}
amendable to standard evidence estimation methods from the Bayesian toolbox, allowing for Bayesian model selection of ERGMs.

In \pkg{Bergm}, the \code{evidence()} function estimates $\widetilde{\pi}(y\mid \models_m)$ in \eqref{eq:cor_pp} using the Chib-Jeliazkov method \citep{chib2} or the power posterior method \citep{fri:pet08, friel6}.

Let us consider the Faux Dixon High School network and specify a higher dimensional model together with its parameter prior distribution. We assume that the model, labeled $\models_{1}$, is almost identical to the model used to generate the simulated data (see \code{?faux.dixon.high}) and includes density (\code{edges}), mutuality (\code{mutual}) and transitivity (\code{gwesp}) effects, plus homophily effects for \code{race}, \code{sex} and \code{grade}; the number of nodes of in-degree 0 and 1 and the number of nodes of out-degree 0 and 1.
We also assume that the normal parameter prior means are centered at 0 except for the \code{edges} parameter which is centered at -5, and that the variance/covariance matrix \code{S.prior1} is diagonal with entries equal to 5.

\begin{CodeChunk} 
\begin{CodeInput} 
R> m1 <- dixon ~ edges + mutual + absdiff("grade") + 
+    nodefactor("race") + nodefactor("grade") + nodefactor("sex") +
+    nodematch("race", diff = TRUE, levels = c("B","O","W")) + 
+    nodematch("grade", diff = TRUE) + 
+    nodematch("sex", diff = FALSE) + 
+    idegree(0:1) + odegree(0:1) + gwesp(0.1,fixed = TRUE)

R> M.prior1 <- c(-5, rep(0, 26))
R> S.prior1 <- diag(5, 27)
\end{CodeInput} 
\end{CodeChunk} 

The \code{evidence()} function is used to carry out MCMC sampling from the posterior distribution~\eqref{eq:cor_pp}.
The total number of iterations per chain is \code{burn.in + main.iters}, with the first \code{burn.in} MCMC draws removed from the posterior sample. 
The covariance structure of the multivariate Normal proposal distribution for the MCMC run can be modified by the argument \code{V.proposal}, which can be easily tuned to achieve a suitable acceptance rate by starting from its default value (1.5). 
The approximate contrastive divergence (CD) estimate is preferred to the approximate maximum likelihood estimator for $\htheta_{MLE,m}$, as the function \code{ergm()} that is called from the \code{ergmAPL()} function may require several minutes to estimate the $\htheta_{MLE,m}$, due to the larger size of the network and the higher dimensionality of the parameter space.
The importance sampling algorithm for estimating the magnitude adjustment constant $C_m>0$ was carried out using \code{ladder = 200} path points, \code{aux.iters= 2500} auxiliary iterations used for drawing the first network from the ERGM likelihood at each iteration, \code{n.aux.draws= 50} auxiliary networks drawn from the ERGM likelihood and \code{aux.thin= 50} auxiliary iterations between each of the \code{n.aux.draws} network draws after the first network is drawn.

\begin{CodeChunk} 
\begin{CodeInput} 
R> cj1 <- evidence(
+    evidence.method = "CJ",
+    formula         = m1,
+    prior.mean      = M.prior1,
+    prior.sigma     = S.prior1,
+    aux.iters       = 2500,
+    n.aux.draws     = 50,
+    aux.thin        = 50,
+    ladder          = 200,
+    V.proposal      = 0.5,
+    burn.in         = 5000,
+    main.iters      = 30000,
+    num.samples     = 25000, 
+    estimate        = "CD",
+    seed            = 1)
\end{CodeInput} 
\end{CodeChunk} 
The posterior summaries can be obtained by using the \code{summary()} function, with some output below omitted for reasons of space.
\begin{CodeChunk}
\begin{CodeInput}
R> summary(cj1)
\end{CodeInput}
\end{CodeChunk}

\begin{CodeChunk}
\begin{CodeOutput}
 Posterior Density Estimate for Model: y ~ edges + mutual + absdiff("grade") + 
	nodefactor("race") + nodefactor("grade") + nodefactor("sex") + 
	nodematch("race", diff = TRUE, levels = c("B", "O", "W")) + 
	nodematch("grade", diff = TRUE) + nodematch("sex", diff = FALSE) + 
	idegree(0:1) + odegree(0:1) + gwesp(0.1, fixed = TRUE) 
 
                                     Mean        SD    Naive SE Time-series SE
theta1 (edges)                -4.90224106 1.4719285 0.009309293     0.08504900
theta2 (mutual)                1.83352189 1.1779171 0.007449802     0.07034189
theta3 (absdiff.grade)        -0.54286164 0.4143574 0.002620626     0.02417282
theta4 (nodefactor.race.H)     0.30639605 1.3374887 0.008459022     0.07841541
theta5 (nodefactor.race.O)     0.65338715 1.0463389 0.006617628     0.06260614
theta6 (nodefactor.race.W)     0.20596298 0.9187085 0.005810423     0.05136997
theta7 (nodefactor.grade.8)   -0.09369201 0.8291035 0.005243711     0.04797846
theta8 (nodefactor.grade.9)    0.18362113 0.7716447 0.004880309     0.04336320
theta9 (nodefactor.grade.10)  -0.24990956 0.8461726 0.005351665     0.05063688
theta10 (nodefactor.grade.11)  0.24427172 0.8186635 0.005177682     0.04864007
theta11 (nodefactor.grade.12) -0.11065898 1.0002607 0.006326204     0.05799441
theta12 (nodefactor.sex.2)    -0.29269027 0.4345187 0.002748138     0.02438447
theta13 (nodematch.race.B)     1.11876543 1.2070394 0.007633987     0.07035913
theta14 (nodematch.race.O)    -0.20702689 2.0337078 0.012862298     0.11337410
theta15 (nodematch.race.W)     0.86330856 1.4077270 0.008903247     0.08121955
theta16 (nodematch.grade.7)    0.20816749 1.7890028 0.011314647     0.10371254
theta17 (nodematch.grade.8)    0.28435516 1.1955435 0.007561281     0.06522766
theta18 (nodematch.grade.9)   -0.48123424 1.1959714 0.007563987     0.06629442
theta19 (nodematch.grade.10)   0.51338594 1.4993214 0.009482541     0.09148370
theta20 (nodematch.grade.11)   0.44266211 1.7318564 0.010953221     0.10032692
theta21 (nodematch.grade.12)   0.17301808 1.5826357 0.010009467     0.09034579
theta22 (nodematch.sex)        0.10512247 0.5721881 0.003618835     0.03180088
theta23 (idegree0)             0.23974076 1.3027694 0.008239437     0.07682043
theta24 (idegree1)             0.03379436 0.7498571 0.004742512     0.04377110
theta25 (odegree0)             0.78620919 1.7803523 0.011259937     0.10648695
theta26 (odegree1)            -0.67732871 0.9424583 0.005960629     0.05609142
theta27 (gwesp.fixed.0.1)      1.20023026 0.3959577 0.002504256     0.02305677

 Acceptance rate: 0.2 
\end{CodeOutput}
\end{CodeChunk}

\begin{figure}[H]
\centering
\includegraphics[width=15cm]{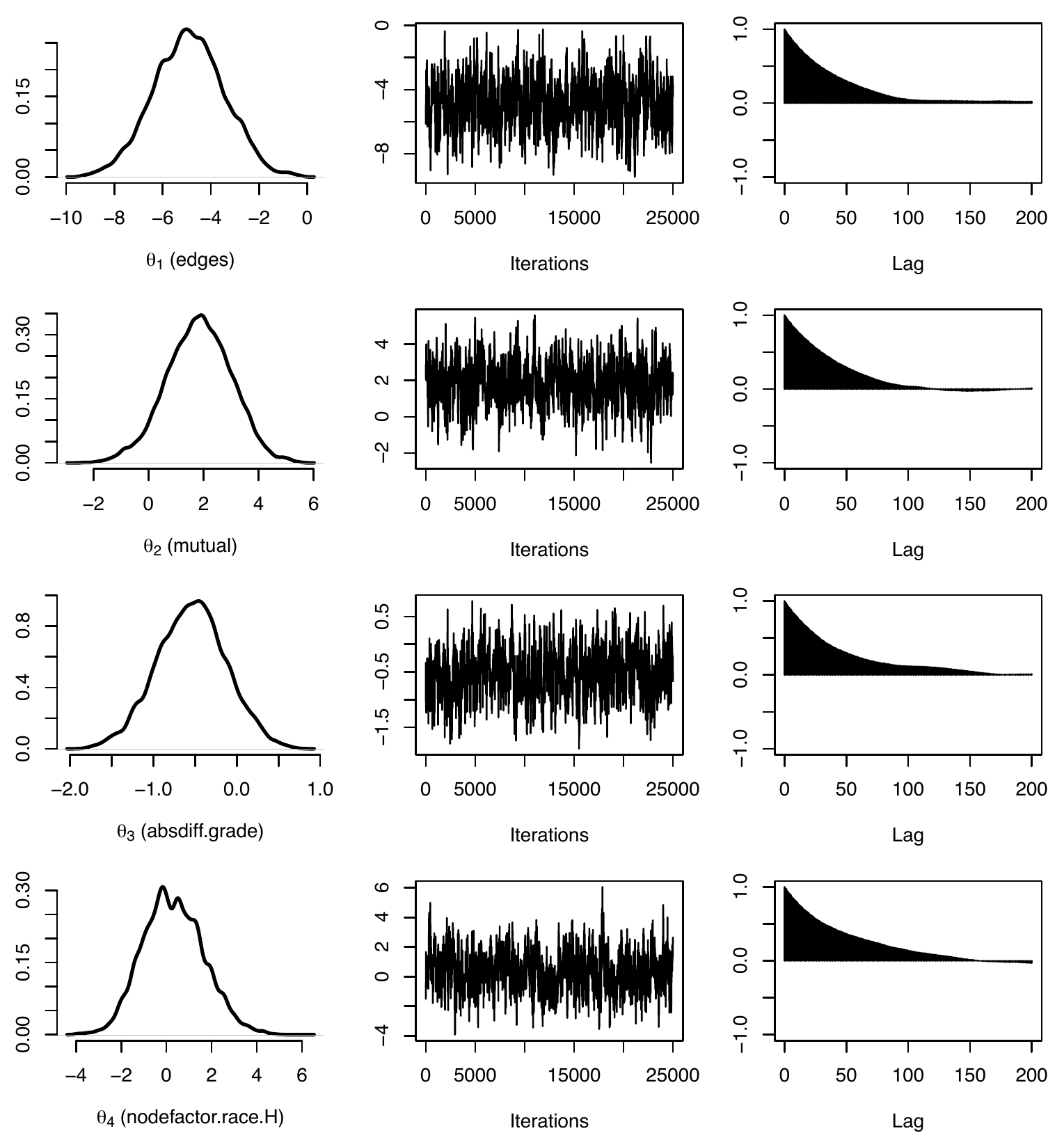}
\caption{MCMC diagnostics for model \code{cj1} (first four parameters only).}
\label{fig:cj1_out}
\end{figure}

Figure \ref{fig:cj1_out} displays the MCMC diagnostic plots produced by the \code{plot()} function for the first four model parameters. The overall acceptance rate is about 20\% and the autocorrelation is negligible after lag 100. The posterior parameter estimates obtained by the \code{evidence()} function can be assessed using the \code{bgof()} function.

\begin{CodeChunk}
\begin{CodeInput}
R> plot(cj1)
\end{CodeInput}
\end{CodeChunk} 

The set of GOF statistic distributions used for the comparison of directed graphs (such as the Dixon network) includes the in-degree distribution, the out-degree distribution, the minimum geodesic distance distribution and the edgewise shared partner distribution. The arguments \code{n.ideg}, \code{n.odeg},  \code{n.dist}, and \code{n.esp} indicate the number of boxplots to plot for each distribution, respectively.

\begin{CodeChunk} 
\begin{CodeInput} 
R> bgof(cj1,  
+    sample.size = 100, 
+    aux.iters   = 5000,
+    n.ideg      = 20, 
+    n.odeg      = 20, 
+    n.dist      = 10, 
+    n.esp       = 7)
\end{CodeInput} 
\end{CodeChunk}

\begin{figure}[H]
\centering
\includegraphics[width=14cm]{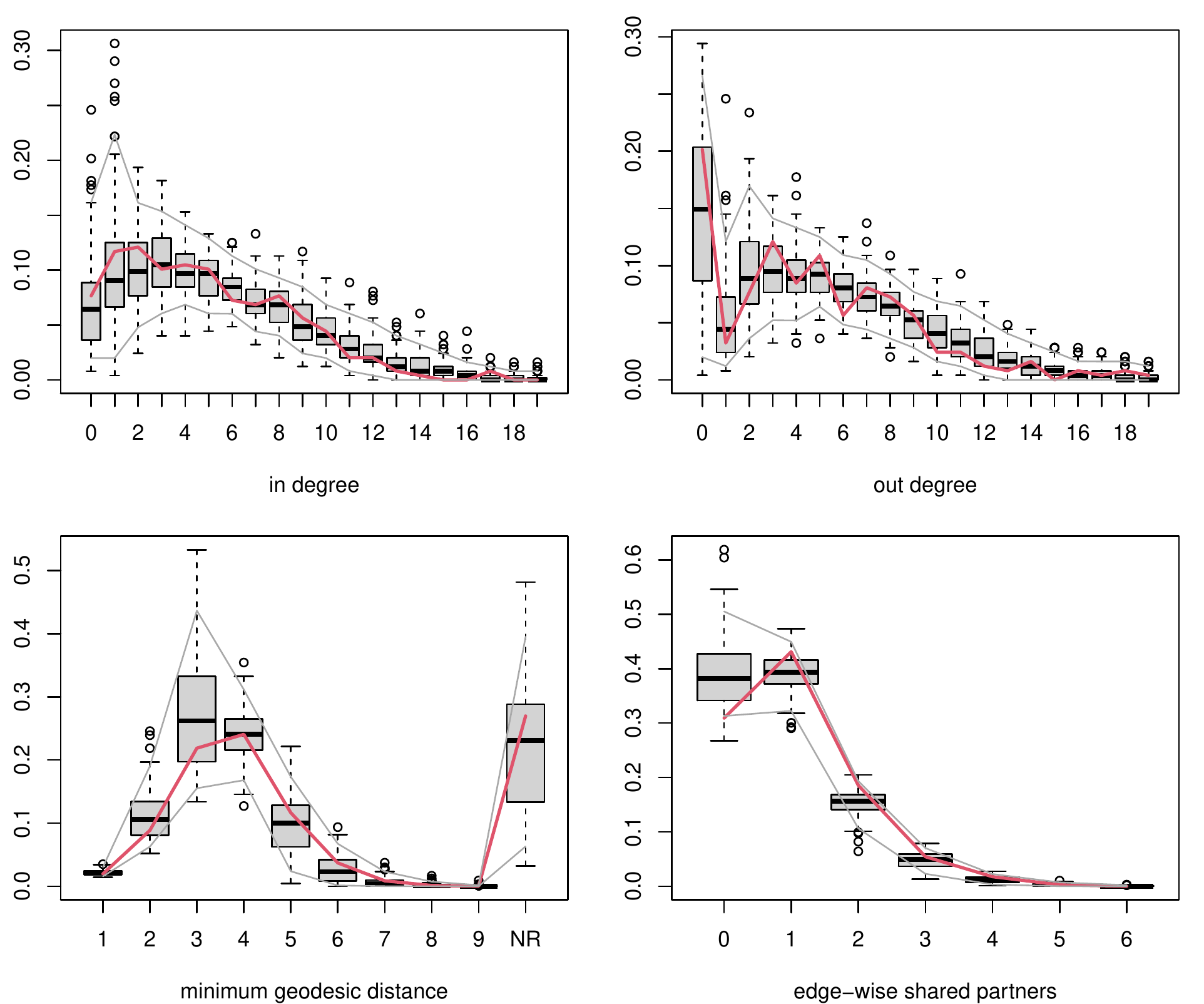}
\caption{Bayesian goodness-of-fit diagnostics for the estimated parameter posterior distribution of model \code{cpp.m0}.}
\label{fig:dixon_gof}
\end{figure}

Figure~\ref{fig:dixon_gof} shows that the \code{cpp.m0} model performs well in reproducing the observed global network properties that are not included in the model. 

The \code{evidence()} function includes an additional feature that allows for estimation of the log-model evidence using the fully adjusted pseudo-likelihood; it is a wrapper for the functions \code{evidenceCJ()} and \code{evidencePP()}, which implement Chib and Jeliazkov's method and the power posteriors method, respectively. For this example, Chib and Jeliazkov's method was selected with the option \code{evidence.method = "CJ"} and \code{num.samples = 25000} MCMC draws were kept for estimating the log-evidence.

We proceed with a model selection example by defining model $\models_{2}$, which assumes that the terms for mutuality and transitivity are removed from $\models_{1}$.

\begin{CodeChunk} 
\begin{CodeInput} 
R> m2 <- dixon ~ edges + absdiff("grade") + 
+    nodefactor("race") + nodefactor("grade") + nodefactor("sex") +
+    nodematch("race", diff = TRUE, levels = c("B","O","W")) + 
+    nodematch("grade",diff = TRUE) + 
+    nodematch("sex",  diff = FALSE) + 
+    idegree(0:1) + odegree(0:1)

R> M.prior2 <- c(-5, rep(0, 24))
R> S.prior2 <- diag(5, 25)

R> cj2 <- evidence(
+    evidence.method = "CJ",
+    formula         = m2,
+    prior.mean      = M.prior2,
+    prior.sigma     = S.prior2,
+    aux.iters       = 2500,
+    n.aux.draws     = 50,
+    aux.thin        = 50,
+    ladder          = 200,
+    V.proposal      = 0.5,
+    burn.in         = 5000,
+    main.iters      = 30000,
+    num.samples     = 25000,
+    estimate        = "CD",
+    seed            = 1)
\end{CodeInput} 
\end{CodeChunk} 

When two models are equally probable a priori, so that $p(\models_{1}) = p(\models_{2}),$ the Bayes Factor is equal to the posterior odds ratio of $\models_{1}$ and $\models_{2}$ in Equation~\eqref{eqn:por}. 
In this example the estimated Bayes Factor provides strong evidence in favour of model $\models_{1}$, as expected (Table \ref{tab:bayes}).
This reveals that the transitivity and mutuality effect are important connectivity features of the observed network but also the homophily effect of \code{race}, \code{sex} and \code{grade} can help explain the complexity of the observed network data. 
Under this setting and the increased model complexity, the CPU time for each implementation of Chib and Jeliazkov's method is of the order of minutes (Table \ref{tab:bayes}). The same estimation would require a few hours using the \code{bergm()} function.

\begin{table}[H]
\caption{Faux Dixon High School - Log evidence  estimates, CPU time in minutes and resulting Bayes Factor (BF) estimate for the models under consideration.}
\vspace{-2.0em}
\begin{center}
\begin{tabular}{lrrr}
\toprule
Model    &  Log evidence estimate & CPU (mins) & $\text{BF}_{12}$\\ 
\hline
$\models_{1}$  & $-38,064.65$ & $2.93$ & $3.68\times 10^{58}$\\
$\models_{2}$  & $-38,199.50$ & $1.13$ & \\
\bottomrule
\end{tabular}
\end{center}
\label{tab:bayes} 
\end{table}

\section{Discussion}
The software package \pkg{Bergm} aims to help researchers and practitioners in two ways. Firstly, it provides a simple, efficient and complete range of tools for conducting Bayesian inference for exponential random graph models. Secondly, \pkg{Bergm} makes available a platform that can be easily customised, extended, and adapted to address different requirements. The software package is under continual maintenance and periodic significant upgrading. Future developments will include uncertainty quantification of the Monte Carlo estimates of the evidence and extensions to weighted networks \citep{cai:gol20}.

The aim of this tutorial is to serve as a useful introduction to the main capabilities of the package as well as some of the algorithms and methods behind it.

\end{document}